\documentclass[%
 aip,
 amsmath,amssymb,
 reprint,%
]{revtex4-1}

\usepackage{graphicx}
\usepackage{dcolumn}
\usepackage{bm}

\usepackage[utf8]{inputenc}
\usepackage[T1]{fontenc}
\usepackage{mathptmx}

\begin{document}

\preprint{AIP/123-QED}

\title{Simultaneous Phase Separation and Pattern Formation  in Chiral Active Mixtures}

\author{Demian Levis}  \email{demian.levis@epfl.ch}
\affiliation{CECAM Centre Europ\'een de Calcul Atomique et Mol\'eculaire, \'Ecole Polytechnique F\'ed\'erale de Lausanne, Batochime, Avenue Forel 2, 1015 Lausanne, Switzerland}
\affiliation{Departament de F\'isica de la Mat\`eria Condensada, Universitat de Barcelona, Mart\'i i Franqu\`es 1, E08028 Barcelona, Spain}
\affiliation{UBICS University of Barcelona Institute of Complex Systems, 
Mart\'i i Franqu\`es 1, E08028 Barcelona, Spain}

\author{Benno Liebchen}%
\affiliation{Institut f\"{u}r Theoretische Physik II: Weiche Materie, Heinrich-Heine-Universit\"{a}t D\"{u}sseldorf, D-40225 D\"{u}sseldorf, Germany}

\date{\today}

\begin{abstract}
Chiral active particles, or self-propelled circle swimmers, from sperm cells to asymmetric Janus colloids, 
form a rich set of patterns, which are different from those seen in linear swimmers.
Such patterns have mainly been explored for identical circle swimmers, while real-world circle swimmers, 
typically possess a frequency distribution. 
Here we show that even the simplest mixture of (velocity-aligning) circle swimmers with two different frequencies, hosts a complex world of superstructures:
The most remarkable example comprises a microflock pattern, formed in one species, while the other species phase separates and forms a macrocluster, coexisting with 
a gas phase. Here, one species microphase-separates and selects a characteristic length scale, whereas the other one macrophase separates and selects a density. 
A second notable example, here occurring in an isotropic system, are patterns comprising two different characteristic length scales, which are controllable via 
frequency and swimming speed of the individual particles.
\end{abstract}

\maketitle

\section{\label{sec:Intro}Introduction}
Chirality,
the property of a structure to be distinguishable from (or not superimposable with) its mirror image, 
plays an important role in all natural sciences. 
In physics, for example, the concept of chirality plays an important role from subatomic scales - for nucleonic mass generation - to astronomical scales - for the formation of galaxies, commonly showing a 
disk-like geometry with bright spiral arms witnessing ongoing star-formation.
In biology, chirality shows up, for instance, in the double-helical structure of DNA, the shape of bacterial flagella or 
the anatomy of flatfish like Halibut. 
Interestingly, in many cases left- and right-handed chiral structures are not equally distributed:
For instance, bacterial flagella and 19 out of 20 amino acids are left-handed, evoking 
questions regarding the origin and possible purpose of the prevalence of a certain handedness upon chiral structures. (Did it emerge before or from life, 
on earth, or does it have an extraterrestrial origin?) 

Chirality also occurs in active matter, comprising self-propelled particles such as microswimmers.  
Here, microswimmers with mirror-symmetric body parts 
swim linearly, whereas those with chiral body shapes (or body parts), show noisy circular trajectories in 2D and helical trajectories in 3D 
\cite{Lowen2016,Friedrich2016}. 
Biological examples of chiral microswimmers include sperm cells \cite{Riedel2005,Yang2008,Elgeti2015} and
{\it E. Coli} bacteria close to walls or interfaces \cite{Maeda1976,DiLuzio2005,Lauga2006}, both featuring chiral body parts determining 
the handedness of their swimming trajectories. Therefore, ensembles of chiral biological microswimmers 
often share the same chirality (monochirality), but show a distribution of rotation frequencies. 
Conversely, man-made synthetic microswimmers, like asymmetric Janus-colloids
\cite{Kummel2013,Wykes2016} or granular microflyers \cite{Scholz2018}
allow to engineer the handedness of a microswimmer on demand (e.g. via 3D printing \cite{Scholz2016}). Thus, also 
polychiral mixtures can be produced, which in principle could be reduced 
to monochiral mixtures using chiral segregation schemes \cite{Mijalkov2013,Chen2015,Levis2018}. 

Besides affecting the trajectories of single active particles, in free space or in crowded environments \cite{Chepi}, chirality can also have a spectacular impact on the collective behavior of circle swimmers.
Specifically, for the single frequency (monochromatic) case, it has been shown that 
circle swimmers with a tendency to align can self-organize into synchronized rotating doublets \cite{Kaiser2013} and large clusters \cite{Denk2016,Liebchen2017}, providing a potential microscopic basis 
for the rotating ring-clusters observed in self-propelled membrane-bound FtsZ-proteins \cite{Loose2014, Denk2016}. 
This class of circle swimmers can also self-organize into a pattern 
of rotating microflocks with a well-defined length-scale which can be controlled by the swimming speed and rotation frequency of the individual microswimmers \cite{Liebchen2017,Levis2018micro},  resembling the patterns seen in ensembles of sperm cells \cite{Riedel2005}.
Circle swimmers with spherical body shapes, which do not align, but sterically repel each other can 
form hyperuniform states \cite{Lei2018} and aggregate in (macro)clusters \cite{Liao2018,Reichhardt2018} which can even
counterrotate with respect to the surrounding gas 
\cite{Liao2018}.

As compared to the monochromatic case, less is known about the patterns emerging in
(aligning) circle swimmers with different frequencies. For such mixtures, 
broadly occurring both in nature and in the world of synthetic microswimmers, 
previous work has focused on synchronization of circle swimmers \cite{Levis2018}, reporting the formation of counterrotating macroclusters as a side result and similar structures discussed in \cite{Ai2018}.
Here, we explore and characterize pattern formation in chiral active mixtures more systematically and provide a state diagram significantly extending the one in \cite{Levis2018}. 
As our key result, we find that this phase diagram comprises a class of unexpected superstructures, occurring generically in a wide domain of parameter space. 
The most interesting example for such a superstructure emerges for a mixture of two species with significantly different intrinsic frequencies: 
they self-organize into a microflock pattern, formed in one species, coexisting with a macrocluster formed by the other species and  
a gas phase. This remarkable pattern unites microphase- and macrophase-separation: one species selects a length scale, the other one a density, both being characteristic, i.e. independent of 
system size. 
(Notice that these superstructures do not emerge
from superimposed patterns formed by each species individually, since each of the species on its own would stay in the disordered phase.)
When both species rotate sufficiently fast, they form a second type of superstructure given by a pattern comprising two characteristic length scales: to form this pattern, 
circle-swimmers also self-sort 
by chirality and form individual microflocks with a species-selective size.
In each case, the length scales involved in the superstructures we report can be controlled by the properties of the individual components of the system (swimming speed and frequency), rather than requiring a more involved 
design of their interactions, as often required to control pattern formation. 
Therefore, mixtures of circle swimmers provide a route to the formation of controllable superstructures, which might serve as a useful design principle to create 
active materials.

\section{\label{sec:Model}Chiral Active Particle Model}
We consider $N$ overdamped circle-swimmers, at positions $\boldsymbol{r}_i(t)$, at time $t$, which 
self-propel with a constant speed $v_0$ in directions $\boldsymbol{n}_i(t)=(\cos\theta_i(t),\,\sin\theta_i(t))$ in a square box of 
linear size $L$ with periodic boundary conditions. The orientation of particle $i$
changes due to an intrinsic frequency $\omega_i$, rotational diffusion and alignment interactions (of strength coefficient $K$) with its neighbors, yielding \cite{Liebchen2017,Levis2018}
\begin{eqnarray}
\dot{\boldsymbol{r}}_i(t)&=& v_0 \boldsymbol{n}_i (t)\label{eq:EOM1} \\
\dot{{\theta}}_i (t)&=& \omega_i +\frac{K}{\pi R^2}\sum_{j\in\partial_i}\sin(\theta_j(t)-\theta_i(t))+\sqrt{2D_r}{\eta}_i (t)\label{eq:EOM2}
\end{eqnarray}
The sum runs over all the neighbors $j$ at a distance less than the interaction radius $R$ to the $i$-th particle, 
$\eta$ represents a Gaussian white noise of zero mean and unit variance and $D_r$ is the rotational diffusion coefficient. 
\\\textbf{State of the art and limiting cases:}
Before discussing pattern formation in mixtures of circle swimmers, let us briefly review what is known for some relevant limiting cases of this model: 
\begin{itemize}
\item (i) Individual circle swimmers: In the absence of interactions ($K=0$), 
each circle-swimmer shows circular Brownian motion with an average radius $v_0/\omega_i$ \cite{vanTeeffelen2008, Sevilla2016}. 
\item (ii) Linear Swimmers: When switching on aligning interactions ($K>0$) but considering non-rotating swimmers $\omega_i=0$, 
particles tend to locally align their swimming direction \cite{Levis2018}. This kind of ferromagnetic, or polar,  
coupling can lead to flocking \cite{VicsekRev}, which occurs when the coupling exceeds a critical strength $K>K_c$ 
and allows a macroscopic fraction of the system to balistically move in a preferred direction. This yields
long-range order in a 2D system with local coupling \cite{Toner1995}.
\item (iii) Single frequency (monochromatic) circle swimmers: In the presence of rotations, the critical coupling strength $K_c$ does not change. 
However, despite having little impact on the transition to flocking itself, rotations (or active torques) dramatically change the collective behavior of polar active particles in the ordered phase.
 For slow rotations, circle-swimmers form a macroscopic rotating cluster which coarsens and scales with the system size at late times, 
whereas faster rotations lead to a pattern of synchronized rotating clusters, or microflocks, with a characteristic size. 
This size scales linearly with the single particle 
radius \cite{Liebchen2017,Levis2018micro} offering a way to control the assembly of chiral active particles. 
\item (iv) No self-propulsion: For several frequencies, but in the absence of self-propulsion ($v_0=0$), eq. \ref{eq:EOM2} reduces to the (noisy) Kuramoto model of phase synchronization \cite{Acebron2005}, and if rotations are also absent, to the  XY model of magnetism \cite{berezinskii1971}, in a 2D  geometric network. 
The two latter models have been largely studied, and it is known that they cannot sustain global (long-range) order in 2D \cite{Daido1988, Mermin1966}. However, remarkably, we now know that self-propulsion ($v_0>0$) 
allows for global synchronization, which can occur e.g. in the form of a mutual flocking phase generalizing the Toner-Tu phase to circle swimmers \cite{Levis2018}.
\end{itemize}
In the following, we focus on pattern formation in mixtures of circle swimmers, which has been far less explored than the above limiting cases. 
For simplicity, we consider two species 
with frequencies $\omega_{1,2}$ which can either have the same sign, representing a monochiral mixture, 
or different signs.
\\\textbf{Units, parameters and simulations:}
We use the interaction range $R$ and the inverse of the rotational diffusion coefficient, $1/D_r$, as length and time units, respectively. 
We define the following dimensionless parameters: (i) mean density per species, 
$\rho_{\alpha}=N_{\alpha}R^2/L^2$ and overall mean density $\rho_0=N R^2/L^2$,
where $N_{\alpha}$ is the number of particles of species $\alpha$, 
with intrinsic frequency $\omega_{\alpha}$; (ii) the reduced frequencies $\Omega_i=\omega_i/D_r$
(iii) the coupling strength $g=K/(4\pi R^2 D_r)$ and (iv) the P\'eclet number $\text{Pe}=v_0/(RD_r)$ (which we fix at $\text{Pe}=2$). 
For simplicity, we focus on the case of equal density per species $\rho_1=\rho_2=\rho_0/2$ 
($\rho_1=\rho_2=\rho_3=\rho_0/3$ in section \ref{sec:Multi}).
To explore pattern formation in mixtures of circle swimmers, we use
Brownian Dynamics simulations of $N=10^3$ up to $N=32\times 10^3$ particles using an Euler integration scheme with a time step $\Delta t=10^{-3}$. 
We then analyze the system at  long-times, after letting it relax for more than $10^4$ times the rotational diffusion time ($tD_r=10^4$) from a random initial configuration.
As shown in \cite{Levis2018}, at the level of a coarse-grained description of the model eq. \ref{eq:EOM1}, \ref{eq:EOM2}, 
the disordered state generically looses stability at $g\rho_0=2$
 (which is robust against excluded volume interactions \cite{Levis2018micro,Aitor}).
Thus, to study pattern formation, we choose $g\rho_0=3$ in the following. 
Note here, that in this case, the density per species is too low to induce pattern formation if it was the only species present; hence, particles
of different species have to cooperate to form patterns.

\section{\label{sec:Sym} Symmetric Mixtures}
\begin{figure}
\includegraphics[scale=0.9]{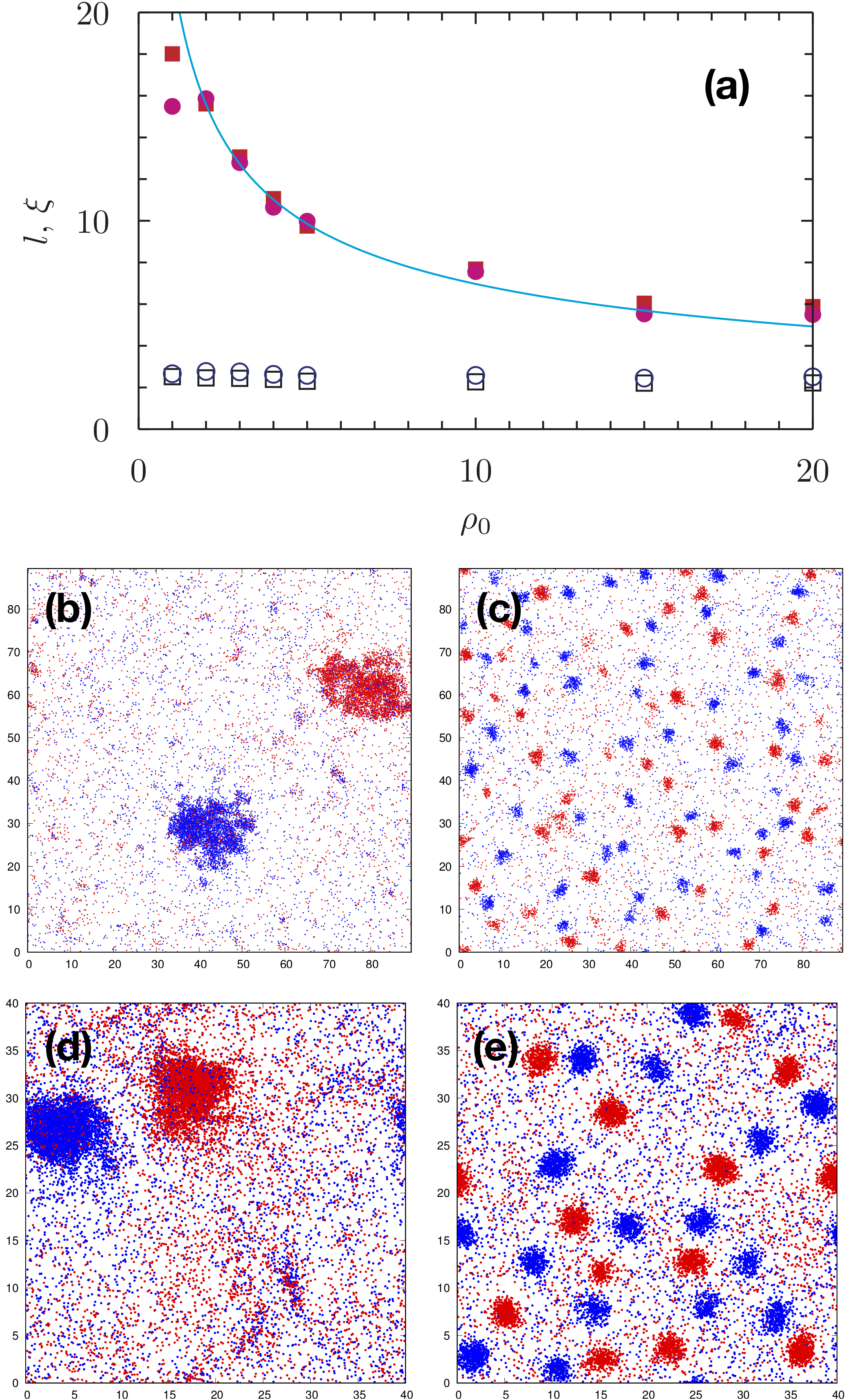}
\caption{\label{fig:xi_vs_rho} (a) Characteristic length scales $l$ (rectangles) and $\xi$  (disks) as a function of  $\rho_0$ at  fixed $g\rho_0=3$ in the macrocluster ($\Omega=0.5$, in filled symbols) 
and microflock ($\Omega=5$, in empty symbols) regime; 
the blue line shows a $~1/\sqrt{\rho_0}$ scaling law (at fixed particle number), i.e. $l,\xi \sim L$, showing 
that macroclusters scale linearly with the system size. 
Configuration snapshots of the system for $\rho_0=2$, $\Omega=0.5$ (b) and $\Omega=5$ (c) and $\rho_0=10$, $\Omega=0.5$ (d) and $\Omega=5$ (e). As the size of the system increases the size of the clusters in the macrocluster regime grows, 
while the number of microflocks in the system increases but keep approximately the same size.  Here we used $N=16\times 10^3$ particles.} \label{fig:rho}
\end{figure}

We first discuss unbiased symmetric mixtures $\Omega_1=-\Omega_2=\Omega$.
\begin{figure}
\includegraphics[scale=0.4]{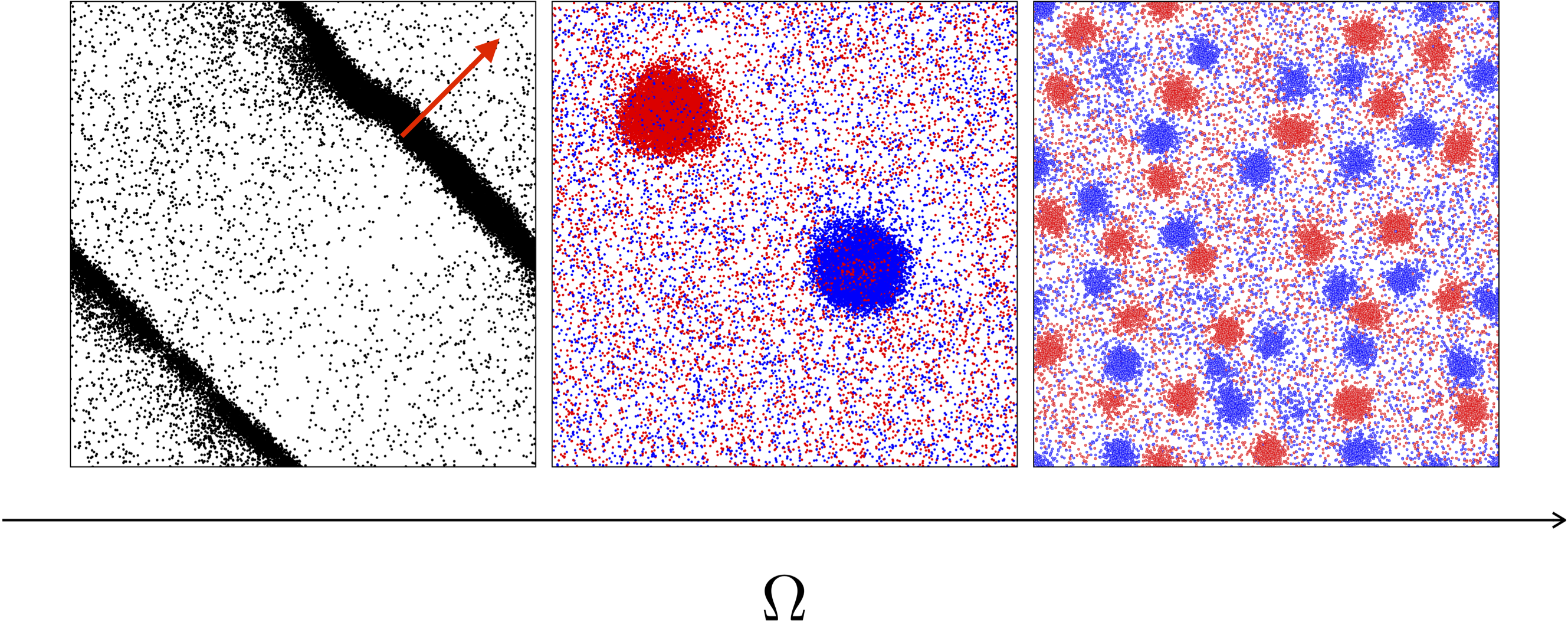}
\caption{\label{fig:sym} Snapshots based on Brownian Dynamics simulations for symmetric mixtures with $\Omega\equiv \Omega_1=-\Omega_2$ 
for increasing rotation frequency: (from left to right) $\Omega=0$, $\Omega=1$ and $\Omega=5$, at fixed $\rho_0=20$ and $N=32\times10^3$. 
The red arrow in the leftmost snapshot represents the average polarization of the particles in the  dense band. Particles rotating at $\Omega$ are represented in blue and the ones rotating at $-\Omega$ in red. Linear swimmers are represented in black. }
\end{figure}
When the coupling is weak ($g\rho_0 \lesssim 2$) the positions and orientations of the particles are randomly distributed, 
leading to a disordered homogeneous gas-like phase (not shown). 
If $g\rho_0 > 2$, the disordered phase looses stability and a new state emerges, which may either be a uniform ordered phase, or 
a pattern. 
\\\textbf{Uniform ordered phase:}
If $g\rho_0 \gg 2$ the system can settle in an ordered 
uniform phase which has been mainly explored for $\Omega=0$ (linear swimmers) 
where it features long-range polar order \cite{Toner1995} and giant density fluctuations \cite{Narayan2007}.
Remarkably, a similar phase, the mutual flocking phase, occurs also for chiral particles of opposite handedness ($\Omega>0$), which cooperate and mutually suppress their circular 
motion, forming two superimposed flocks at a relative angle to each other \cite{Levis2018}.
\\\textbf{Patterns:}
If $g\rho_0 \gtrsim 2$ but not too large, the system forms patterns, comprising high density structures of polarly ordered particles which coexist with a 
disordered low density gas. We explore these patterns in the following: 
(i) For $\Omega=0$ (linear swimmers) the dense structures appear in the form of  
traveling bands \cite{Chate2008}, as shown in the leftmost snapshot Fig. \ref{fig:sym} and discussed in the previous section. 
(ii) If $0\ll \Omega \lesssim 1$ 
the onset of flocking is accompanied by the spatial segregation of particles by their chirality. Following segregation,  
chiral particles form polarly ordered rotating clusters which are roughly spherical (see Fig. \ref{fig:sym}), and coarsen as time proceeds. 
Notably, each cluster contains some particles of opposite chirality, which therefore 
rotate with a frequency opposite to their intrinsic one. 
The central panel of Fig. \ref{fig:sym} shows such clusters at late times, which 
coexist with a disordered, low density background comprising particles of both species.
Following their size and shape, we call them rotating macroclusters, or simply \emph{macrodrops}.  
(iii) For $\Omega \gtrsim 1$ (e.g. $\Omega=5$, rightmost panel Fig. \ref{fig:sym}), circle swimmers self-sort by chirality, as in case (ii), but 
self-organize into a different state:  a 
pattern of chiral rotating clusters, with a characteristic length-scale. The emergence of such a \emph{microflock} pattern is 
associated with a short-wave length instability of the homogeneous flocking state, 
allowing to predict the size of the microflocks in the single-species case \cite{Liebchen2017}. 
(For the present bichromatic case, a stability analysis of the mutual flocking phase would be needed to characterize the onset
and length scales of microflock patterns, which is well-beyond the scope of the present article.)
\\\textbf{Characterization of patterns: Macrodrops and Microflock Patterns}
To characterize the observed patterns, we analyze the size of the clusters in different regimes. We associate a characteristic length 
scale to the clusters based on the analysis of the pair correlation function
\begin{equation}
N\rho_0 G(r)=\langle \delta(|\boldsymbol{r}-\boldsymbol{r}_j+\boldsymbol{r}_i|)\rangle
\end{equation}
and orientational self-correlation function
\begin{equation}
C(r)=\langle \boldsymbol{n}_i\cdot \boldsymbol{n}_j\rangle_{r}
\end{equation}
where $\langle. \rangle_{r}$ denotes an average over all the pairs of particles at distance $r$.
We thus define two length scales for such rotating clusters, representing a density- and an orientational- correlation length: we define $l$ via 
the criterion $G(l)=1$; and $\xi$ via $C(\xi)=e^{-1}$. 
In Fig. \ref{fig:rho} (a) we show both length scales in the slow-rotating, macrodrop regime ($\Omega=0.5$), and in the fast-rotating, microflock one  ($\Omega=5$) 
as a function of $\rho_0$ at fixed $g\rho_0=3$ (above the onset of flocking) and  $N=16.10^3$. In each case $l,\xi$ are very similar to each other. 
Here, for $\Omega=5$ (microflock patterns) both length scales are independent of $\rho_0$ when keeping $g\rho_0$ constant. 
(These length scales have been recorded 'at late times' in the simulations, here, at  $tD_r=25.10^4$; for a discussion about a possible coarsening of microflocks on timescales
beyond those involved in the coarsening of the macrodroplets, see \cite{Liebchen2017}.)
Conversely, for $\Omega=0.5$,  the size of the macrodrop clearly decreases as the density increases, in a way which is 
consistent with a $l,\xi\sim1/\sqrt{\rho_0}\sim L$ scaling (at fixed particle number),
as expected for systems undergoing phase separation. As further evidenced by the snapshots Fig. \ref{fig:rho} (b)-(e), 
the size of the macrodrops reduces when decreasing the system size (see (b), (d)), while it is the number of microflocks which 
increases when the density is reduced and not their size. In the macrodrop regime, for a given value of $g\rho_0$, Pe and $\Omega$, 
the system selects a density (the one of the macrodrops), while in the microflock regime the system selects a length scale. 


\section{Monochiral Mixtures}
\begin{figure}
\includegraphics[scale=1]{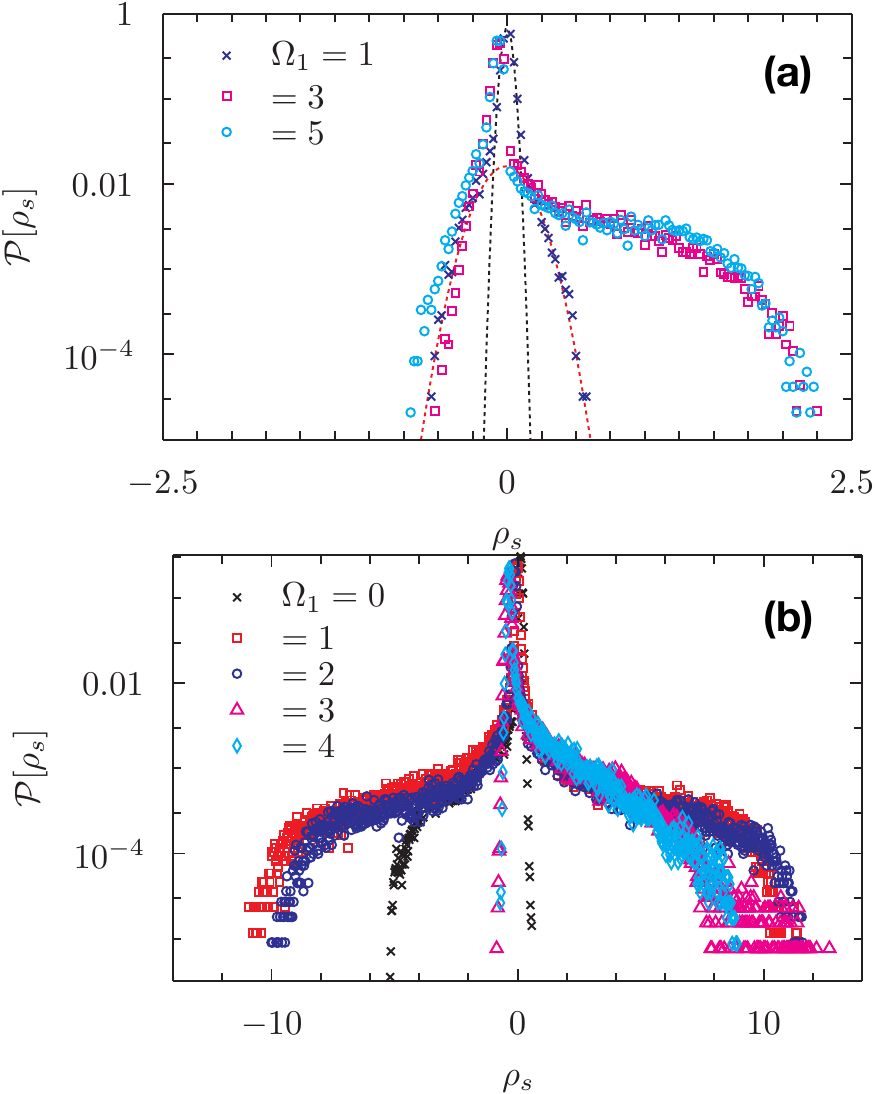}
\caption{
Distribution of the local segregation factor $\mathcal{P}(\rho_s)$ with $\rho_s=(\rho_1-\rho_2)/\rho_0$ for several values of $\Omega_1$ (see key) for mono-chiral mixtures 
(fixed $\Omega_2=1$) (a) and for bi-chiral mixtures ($\Omega_2=-1$) (b). 
The two dotted lines in the top panel correspond to two Gaussian distributions centered in zero with different variance. } \label{fig:pdf}
\end{figure}

We now consider mixtures of swimmers of the same chirality ($\Omega_1\times \Omega_2>0$), which do not fully segregate. 
To see this, we first quantify 'segregation'. 
We compute the local density field of particles of species 1, $\rho_1$, and species 2, $\rho_2$. We then analyze the probability distribution $\mathcal{P}$ of their difference $\rho_s=(\rho_1-\rho_2)/\rho_0$. 
A region with an excess of particles of species $1$ will be characterized by a peak- or shoulder of $\mathcal{P}$ at positive $\rho_s$, 
whereas peaks- or shoulders in $\mathcal{P}$ at negative $\rho_s$ stand for regions with an excess of particles of species $2$. 
Thus, $\mathcal{P}$ allows us to quantify deviations from a homogeneous mixing of circle-swimmers. 
Representative examples of $\mathcal{P}[\rho_s]$ for several values of $\Omega_1$ at fixed $\Omega_2=1$ 
are shown in Fig. \ref{fig:pdf} (a). 
For the case $\Omega_1=\Omega_2$ (single species case), 
$\mathcal{P}[\rho_s]$ features a narrow Gaussian distribution around zero, stemming from particles in an incoherent gas-like state, and a broader Gaussian tail stemming 
from particles in a denser region (a macrodrop, not shown).  
As soon as $\Omega_1>\Omega_2$, the distributions 
become non-symmetric and develops a tail at values of $\rho_s>0$. Such a distribution signifies dense structures with an excess of frequency-$\Omega_1$-particles, 
whereas the uniform background mainly contains $\Omega_2$-particles. (A configuration snapshot of such a state  is shown in Fig. \ref{fig:phd} (d)). 
Thus, for monochiral mixtures, 
fast-rotating particles dominate the 
formation of the dense structures (i.e. of the pattern) while slower ones are partly relegated to the low density regions. 
(This is consistent with the fact that, at the level of field equations\cite{Liebchen2017} for a single-species, 
the growth rate of the microflock-instability increases with the frequency.)
For slower rotations, where the field equations for the single-species case do not show a (short-wavelength) microflock instability
, but a long-wavelength instability\cite{Liebchen2017}, 
this behaviour is less pronounced, and we observe a 
rotating macrocluster continaing a mixture of 
circle swimmers of both frequencies (Fig. \ref{fig:phd}b) - we refer to this case as mixing.

\section{\label{sec:2species} Simultaneity of patterns and macroclusters in generic mixtures}
Conversely to monochiral mixtures, where pattern formation is dominated by the faster rotating species, in symmetric mixtures both species of course behave equally when forming patterns. 
Here, we focus on circle swimmers with opposite handedness but different frequency ($\Omega_1 \neq -\Omega_2$)
and show that they can generate remarkable patterns featuring two characteristic length scales. 

To see this, let us first discuss the segregation behavior of non-symmetric bichiral mixtures (Fig.~\ref{fig:pdf}b).
Conversely to the discussed mixtures where $\mathcal{P}(\rho_s)$ shows only one shoulder (Fig. \ref{fig:pdf}a), 
remarkably, particles of opposite chirality can create two (asymmetric) shoulders in $\mathcal{P}(\rho_s)$ (Fig. \ref{fig:pdf}b).
This leads to a rich set of possible patterns. 
For $|\Omega_1|<|\Omega_2|$ we find that particles of species 2 
can self-organize into dense structures, while particles of species 1 remain rather uniformly distributed, as represented  by the negative tail in 
the distribution for $\Omega_1=0$ Fig. \ref{fig:pdf} (b).
For   $|\Omega_1|=|\Omega_2|$, $\mathcal{P}[\rho_s]$ is symmetric  with large tails at both $\rho_s>0$ and  $\rho_s<0$, indicating the  chiral sorting of particles into dense, counterrotating clusters of same density and size. 
 For $|\Omega_1|>|\Omega_2|$ the distribution is non-symmetric but features a broader tail at $\rho_s>0$, 
corresponding to a tendency of the system to generate dense structures made by a larger fraction of particles of species 1. A configuration snapshot of the system in this case is shown in Fig. \ref{fig:phd} (e): high-frequency swimmers form microflocks while lower-frequency ones form a coexisting single macrocluster. This case represents an example of a state where one species  phase separates (the size of the macrocluster scales with the system size) and the other species forms a pattern with a characteristic length scale. 
For a higher frequency dispersion [see snapshot Fig. \ref{fig:phd} (f)], particles of species 1 quickly form microflocks, leaving no room for slower particles of species 2 to aggregate, as the peak of $\mathcal{P}[\rho_s]$ for $\Omega_1=3$ and $\Omega_1=4$ at small negative values of $\rho_s$ and the tail at large positive values of $\rho_s$ shows [see Fig. \ref{fig:pdf} (b)]. 
In this case we therefore observe a pattern occurring for one species, whereas the 
second species is in the uniform gas phase. 
Finally, for  $|\Omega_2|>2$,  microflock patterns can occur in both species, as the snapshot Fig. \ref{fig:pdf} (g) shows; here, remarkably,  the resulting pattern comprises two different length scales. 
(Note here, that the observed patterns can not be viewed as a simple superposition of patterns formed by both species individually; instead, at $g\rho_0=3/2<2$, each of the species on its own would be in the disordered phase.) 

\begin{figure}
\includegraphics[scale=1]{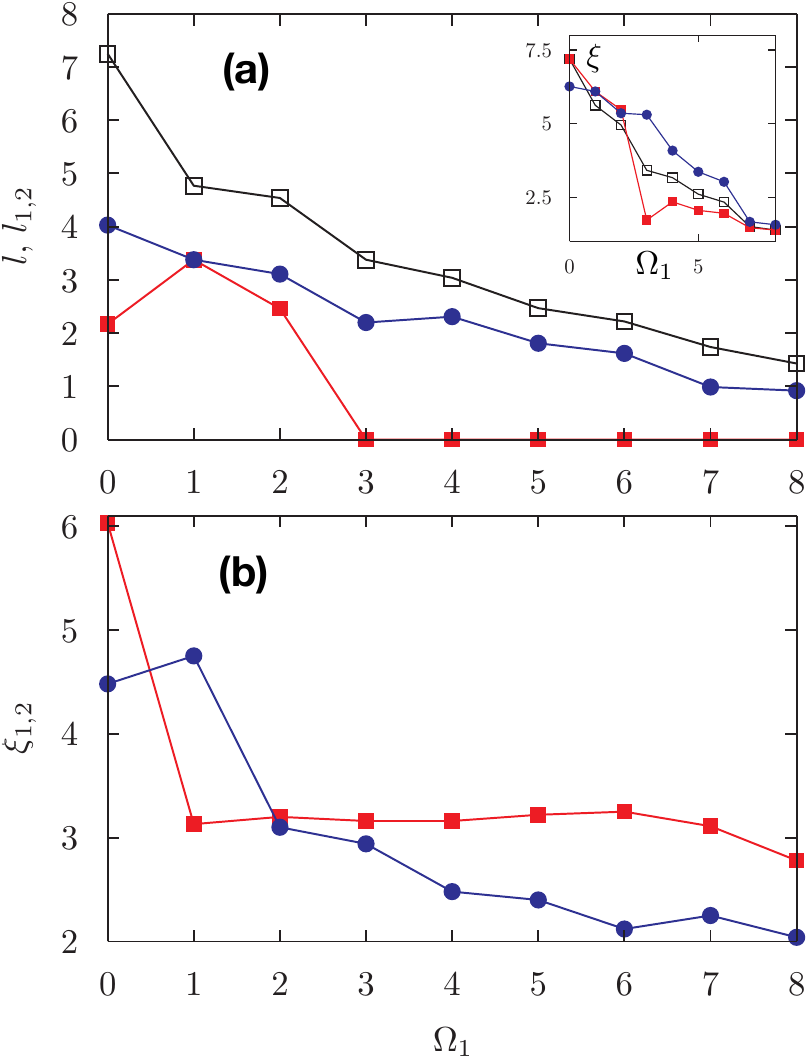}
\caption{ (a): Length-scales associated to particle clustering, $l$ (in empty symbols), $l_1$ (in blue) and $l_2$ (in red), as a function of $\Omega_1$ at fixed $\Omega_2=-1$. The inset shows the correlation length $\xi$ (empty symbols), $\xi_1$ (in blue) and $\xi_2$ (in red).
(b): $\xi_1$ (in blue) and $\xi_2$ (in red), as a function of $\Omega_1$ at fixed $\Omega_2=-2$.  }\label{fig:length}
\end{figure}

To characterize these patterns, we
generalize the definition used for symmetric mixtures to define length scales $l_\alpha,\xi_\alpha$ ($\alpha\in \{1,2\}$), 
associated with the density and orientation correlations for each species. 
We define $l_\alpha$ via $G_\alpha(l_\alpha)=1$, where 
$G(r)$ is the pair correlation function, which is in turn defined via 
$N\rho_0 G_{\alpha}(r)=\langle \delta(|\boldsymbol{r}-\boldsymbol{r}^{\alpha}_j+\boldsymbol{r}^{\alpha}_i|)\rangle$, 
where averages are performed over the particles, indexed with $i,j$. 
From the partial orientational 
correlation function $C_\alpha(r)$, defined via
$C^{\alpha}(r)=\langle \boldsymbol{n}^{\alpha}_i\boldsymbol{n}^{\alpha}_j\rangle$, we define $\xi_\alpha$ via $C(\xi_{\alpha})=1/e$. 
In Fig. \ref{fig:length}, we show these length scales as a function of $\Omega_1$ at fixed $\Omega_2=-1$ (a) and $\Omega_2=-2$ (b). 
In both cases, as $\Omega_1$ increases, the characteristic length scale decreases, i.e. faster rotations induce smaller structures. 
For $\Omega_2=-1$ [see Fig. \ref{fig:length} (a)], $l_1$ and $\xi_1$ decrease roughly linearly with $\Omega_1$, while the length scales associated to particles of species 2 
saturate at some comparatively small values ($l_2=0$ and $\xi_2\approx2$) for $\Omega_1\geq 3$, as expected from the previous inspection of the distribution functions Fig \ref{fig:pdf} (b). 
In this regime, i.e. for large enough $\Omega_1$, particles of species 2 are not able to form segregated dense structures, such that the pattern 
shows a single characteristic length scale. 
Interestingly, the situation changes for larger $\Omega_2$, providing a rich pattern formation scenario.  
Here, as shown in  Fig. \ref{fig:length} (b), two finite length scales associated to particles of species 1 and 2 can coexist. 
For $\Omega_1\geq 1$, particles of species 2 form clusters of a given size $\xi_2\approx 3$, while, by increasing $\Omega_1$, particles of species 1 form structures of smaller size. 
In this regime, the emerging patterns feature two characteristic length scales (see snapshot Fig. \ref{fig:phd} (g)).
 
\section{\label{sec:2species} Phase Diagram}
\begin{figure*}
\includegraphics[scale=0.45]{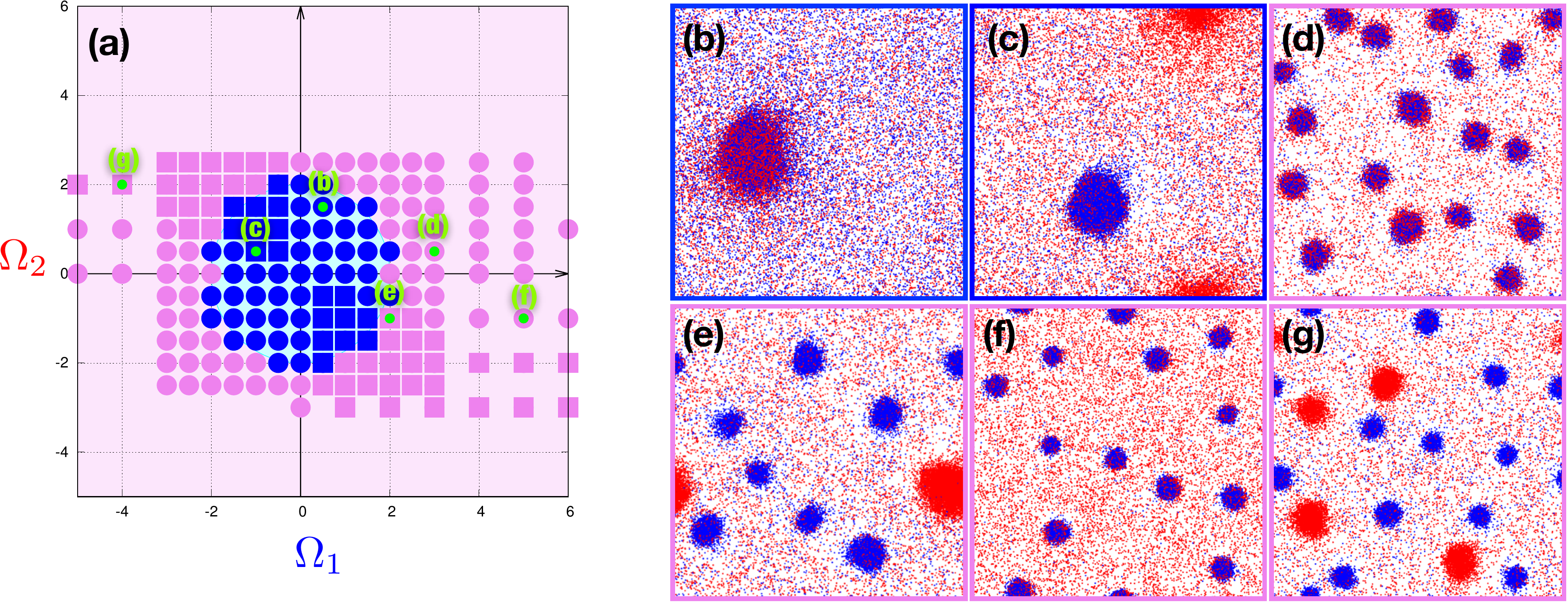}
\caption{\label{fig:phd} State diagram of two-species mixtures of aligning circle-swimmers with rotation frequency $\Omega_1$ and $\Omega_2$ (a), 
together with representative snapshots at different points in the $(\Omega_1,\Omega_2)$ plane: (b): $(0.5,1.5)$; (c): $(-1,0.5)$; (d): $(3, 0.5)$; (e): $(2,-1)$; (f): $(5;-1)$; (g): $(-4;2)$ 
(their location is indicated in the phase diagram (a) by green dots). Particles rotating at frequency $\Omega_1$ are represented in blue, while particles at frequency $\Omega_2$  are in red. 
Blue symbols in (a) denote  macrocluster states, pink ones microflock states. Square symbols correspond to states where particles of opposite chirality segregate into different dense structures. 
The remaining symbols correspond to states where species segregation, if any, does not lead to the formation of species-specific clusters. Note that both macroclusters and microflock patterns of 
opposite chirality can appear, as shown in (g) and (c), respectively (and represented by blue and violet squares in the phase diagram).  
}
\end{figure*}

To provide an overview of the possible patterns seen in mixtures of circle swimmers, we now 
summarize our findings in the state diagram Fig. \ref{fig:phd}. Each symbol in the  $(\Omega_1-\Omega_2)$-plane 
corresponds to a simulation. The pink area shows parameter regimes leading to  microflock patterns while the blue one corresponds to the rotating macrodrop regime. 
Here, rectangular symbols represent 
states where swimmers of opposite chirality segregate into distinct dense structures,  which can be either microflocks (giving rise to a  two-length-scale pattern) or macrodrops. 
We distinguish five regimes, which we link to the snapshots discussed earlier. 

For  $\Omega_1^2+\Omega_2^2 \lesssim 2$, the system forms macroclusters. Here, we can either have (i) a single macrocluster containing a mixture of circle swimmers, which occurs in the monochiral case 
(see Fig. \ref{fig:phd} (b)) or by (ii) the formation of two macroclusters of opposite handedness (segregation), 
which generically feature different densities (see Fig. \ref{fig:phd} (c)). 
For  $\Omega_1^2+\Omega_2^2  \gtrsim 2$ microflock patterns emerge. Here, as discussed in the previous section, we can distinguish three different regimes, characterized by:  
(iii) the formation of microflocks made of a larger fraction of fast-rotating particles (see Fig. \ref{fig:phd} (d) and movie 1 in the SM \cite{SM} for the mono-chiral case, and Fig. \ref{fig:phd} (f) and movie 3 \cite{SM} for the bi-chiral one); 
(iv) the simultaneous formation of microflocks and macrodrops composed of swimmers of opposite chirality (see Fig. \ref{fig:phd} (e) and movie 2  \cite{SM}); (v) the formation of microflocks of different size and handedness 
(see Fig. \ref{fig:phd} (g)). The pattern in regimes (iii) and (iv) is characterized by a single length-scale while, strikingly, a two-length-scale pattern emerges in regime (v).

\section{\label{sec:Multi} Multi-species mixtures}
Finally, we briefly comment on generalizations (i) to three species
rotating at frequencies $\Omega_1$, $\Omega_2$ and $\Omega_3$ with $\rho_1=\rho_2=\rho_3$ 
and (ii) to 
a continuous distribution of frequencies where each swimmers' frequency $\Omega_i$ is picked from a uniform distribution $U(\Omega_a;\Omega_b)$. 

\begin{figure}
\includegraphics[scale=0.9]{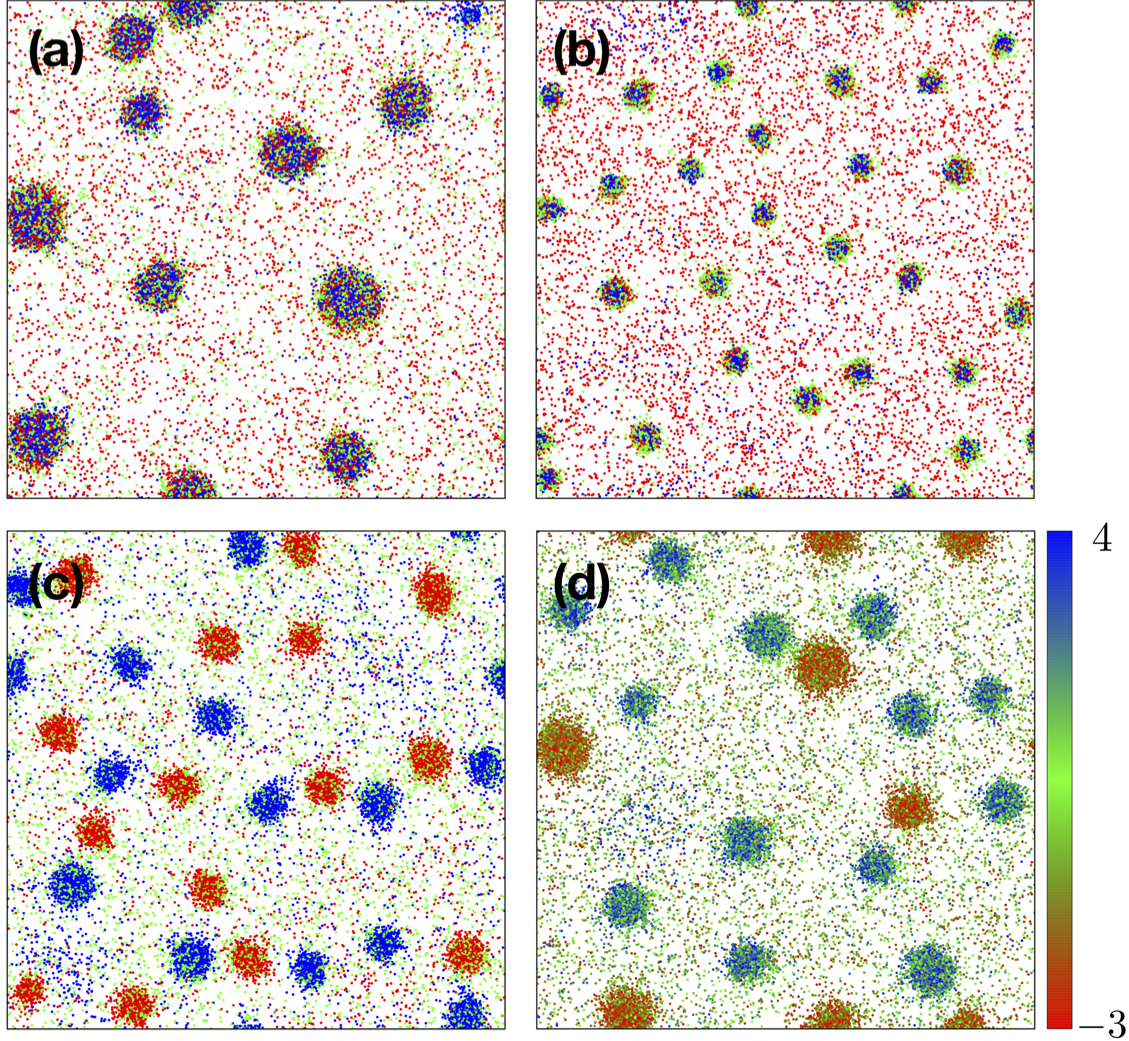}
\caption{\label{fig:3sp} Late-time snapshots for three-species mixtures with frequency $(\Omega_1,\Omega_2, \Omega_3)$ $=(0,2,4)$ (a) , $(-2,4,8)$ (b) and $(-3,0,3)$ (c) and for a continuous mixture of swimmers with uniformly distributed intrinsic frequencies with $\Omega_a=-3$ and $\Omega_b=4$ (d). Particles of each species are colored according to their frequency:  $\Omega_1$-particles are in red, $\Omega_2$ in green and $\Omega_3$ in blue and using a the color code shown in (d) in the continuous polychromatic case. }\label{fig:3sp}
\end{figure}

We show in  Fig. \ref{fig:3sp} configuration snapshots for four cases, using both mono-chiral and bi-chiral mixtures. 
As expected from our discussion above, fast-rotating particles lead to the formation of dense structures, while slowly-rotating ones largely remain in the disordered background. 
The size of the microflock patterns is smaller for larger frequencies (see Fig. \ref{fig:3sp} (b)). For a large enough frequency difference between particles of opposite chirality, 
microflocks of opposite chirality emerge (see Fig. \ref{fig:3sp} (c-d)),  giving rise to, eventually, a pattern characterized by two different length scales [Fig. \ref{fig:3sp} (d)]. 



\section{\label{sec:Conclusion}Conclusions}
We have shown that mixtures of circle swimmers with a tendency to align, show a spectrum of unusual structure formation scenarios. 
While bichiral mixtures tend to spatially segregate and hence self-sort by their chirality, circle swimmers of the same chirality 
can cooperatively form dense clusters, i.e. they mix. Within both regimes, segregation and mixing, chiral active particles can either form 
macroclusters (or macrodrops) with a size scaling with the system size or microflock patterns with a characteristic self-limited size. 
One particularly interesting scenario occurs when one species rotates much faster and opposite to the other one: 
here, a phase-separating macrocluster and a microflock pattern can simultaneously exist, albeit none of the two species would form structures on its own. 
Finally, also the case where both species rotate sufficiently fast in opposite directions is remarkable: 
here, they form a pattern of clusters comprising two length scales
which can be individually controlled by the frequency or self-propulsion velocity of the particles of each species.


\nocite{*}
\bibliography{literature}

\begin{thebibliography}{39}%
\makeatletter
\providecommand \@ifxundefined [1]{%
 \@ifx{#1\undefined}
}%
\providecommand \@ifnum [1]{%
 \ifnum #1\expandafter \@firstoftwo
 \else \expandafter \@secondoftwo
 \fi
}%
\providecommand \@ifx [1]{%
 \ifx #1\expandafter \@firstoftwo
 \else \expandafter \@secondoftwo
 \fi
}%
\providecommand \natexlab [1]{#1}%
\providecommand \enquote  [1]{``#1''}%
\providecommand \bibnamefont  [1]{#1}%
\providecommand \bibfnamefont [1]{#1}%
\providecommand \citenamefont [1]{#1}%
\providecommand \href@noop [0]{\@secondoftwo}%
\providecommand \href [0]{\begingroup \@sanitize@url \@href}%
\providecommand \@href[1]{\@@startlink{#1}\@@href}%
\providecommand \@@href[1]{\endgroup#1\@@endlink}%
\providecommand \@sanitize@url [0]{\catcode `\\12\catcode `\$12\catcode
  `\&12\catcode `\#12\catcode `\^12\catcode `\_12\catcode `\%12\relax}%
\providecommand \@@startlink[1]{}%
\providecommand \@@endlink[0]{}%
\providecommand \url  [0]{\begingroup\@sanitize@url \@url }%
\providecommand \@url [1]{\endgroup\@href {#1}{\urlprefix }}%
\providecommand \urlprefix  [0]{URL }%
\providecommand \Eprint [0]{\href }%
\providecommand \doibase [0]{http://dx.doi.org/}%
\providecommand \selectlanguage [0]{\@gobble}%
\providecommand \bibinfo  [0]{\@secondoftwo}%
\providecommand \bibfield  [0]{\@secondoftwo}%
\providecommand \translation [1]{[#1]}%
\providecommand \BibitemOpen [0]{}%
\providecommand \bibitemStop [0]{}%
\providecommand \bibitemNoStop [0]{.\EOS\space}%
\providecommand \EOS [0]{\spacefactor3000\relax}%
\providecommand \BibitemShut  [1]{\csname bibitem#1\endcsname}%
\let\auto@bib@innerbib\@empty
\bibitem [{\citenamefont {L{\"o}wen}(2016)}]{Lowen2016}%
  \BibitemOpen
  \bibfield  {author} {\bibinfo {author} {\bibfnamefont {H.}~\bibnamefont
  {L{\"o}wen}},\ }\bibfield  {title} {\enquote {\bibinfo {title} {Chirality in
  microswimmer motion: From circle swimmers to active turbulence},}\
  }\href@noop {} {\bibfield  {journal} {\bibinfo  {journal} {Eur. Phys. J.
  Spec. Top.}\ }\textbf {\bibinfo {volume} {225}},\ \bibinfo {pages} {2319}
  (\bibinfo {year} {2016})}\BibitemShut {NoStop}%
\bibitem [{\citenamefont {Friedrich}(2016)}]{Friedrich2016}%
  \BibitemOpen
  \bibfield  {author} {\bibinfo {author} {\bibfnamefont {B.}~\bibnamefont
  {Friedrich}},\ }\bibfield  {title} {\enquote {\bibinfo {title} {Hydrodynamic
  synchronization of flagellar oscillators},}\ }\href@noop {} {\bibfield
  {journal} {\bibinfo  {journal} {Eur. Phys. J. Spec. Top.}\ }\textbf {\bibinfo
  {volume} {225}},\ \bibinfo {pages} {2353} (\bibinfo {year}
  {2016})}\BibitemShut {NoStop}%
\bibitem [{\citenamefont {Riedel}, \citenamefont {Kruse},\ and\ \citenamefont
  {Howard}(2005)}]{Riedel2005}%
  \BibitemOpen
  \bibfield  {author} {\bibinfo {author} {\bibfnamefont {I.~H.}\ \bibnamefont
  {Riedel}}, \bibinfo {author} {\bibfnamefont {K.}~\bibnamefont {Kruse}}, \
  and\ \bibinfo {author} {\bibfnamefont {J.}~\bibnamefont {Howard}},\
  }\bibfield  {title} {\enquote {\bibinfo {title} {A self-organized vortex
  array of hydrodynamically entrained sperm cells},}\ }\href@noop {} {\bibfield
   {journal} {\bibinfo  {journal} {Science}\ }\textbf {\bibinfo {volume}
  {309}},\ \bibinfo {pages} {300} (\bibinfo {year} {2005})}\BibitemShut
  {NoStop}%
\bibitem [{\citenamefont {Yang}, \citenamefont {Elgeti},\ and\ \citenamefont
  {Gompper}(2008)}]{Yang2008}%
  \BibitemOpen
  \bibfield  {author} {\bibinfo {author} {\bibfnamefont {Y.}~\bibnamefont
  {Yang}}, \bibinfo {author} {\bibfnamefont {J.}~\bibnamefont {Elgeti}}, \ and\
  \bibinfo {author} {\bibfnamefont {G.}~\bibnamefont {Gompper}},\ }\bibfield
  {title} {\enquote {\bibinfo {title} {Cooperation of sperm in two dimensions:
  synchronization, attraction, and aggregation through hydrodynamic
  interactions},}\ }\href@noop {} {\bibfield  {journal} {\bibinfo  {journal}
  {Phys. Rev. E}\ }\textbf {\bibinfo {volume} {78}},\ \bibinfo {pages} {061903}
  (\bibinfo {year} {2008})}\BibitemShut {NoStop}%
\bibitem [{\citenamefont {Elgeti}, \citenamefont {Winkler},\ and\ \citenamefont
  {Gompper}(2015)}]{Elgeti2015}%
  \BibitemOpen
  \bibfield  {author} {\bibinfo {author} {\bibfnamefont {J.}~\bibnamefont
  {Elgeti}}, \bibinfo {author} {\bibfnamefont {R.~G.}\ \bibnamefont {Winkler}},
  \ and\ \bibinfo {author} {\bibfnamefont {G.}~\bibnamefont {Gompper}},\
  }\bibfield  {title} {\enquote {\bibinfo {title} {Physics of microswimmers -
  single particle motion and collective behavior: a review},}\ }\href@noop {}
  {\bibfield  {journal} {\bibinfo  {journal} {Rep. Prog. Phys.}\ }\textbf
  {\bibinfo {volume} {78}},\ \bibinfo {pages} {056601} (\bibinfo {year}
  {2015})}\BibitemShut {NoStop}%
\bibitem [{\citenamefont {Maeda}\ \emph {et~al.}(1976)\citenamefont {Maeda},
  \citenamefont {Imae}, \citenamefont {Shioi},\ and\ \citenamefont
  {Oosawa}}]{Maeda1976}%
  \BibitemOpen
  \bibfield  {author} {\bibinfo {author} {\bibfnamefont {K.}~\bibnamefont
  {Maeda}}, \bibinfo {author} {\bibfnamefont {Y.}~\bibnamefont {Imae}},
  \bibinfo {author} {\bibfnamefont {J.-I.}\ \bibnamefont {Shioi}}, \ and\
  \bibinfo {author} {\bibfnamefont {F.}~\bibnamefont {Oosawa}},\ }\bibfield
  {title} {\enquote {\bibinfo {title} {Effect of temperature on motility and
  chemotaxis of escherichia coli.}}\ }\href@noop {} {\bibfield  {journal}
  {\bibinfo  {journal} {J. Bacteriol.}\ }\textbf {\bibinfo {volume} {127}},\
  \bibinfo {pages} {1039--1046} (\bibinfo {year} {1976})}\BibitemShut {NoStop}%
\bibitem [{\citenamefont {DiLuzio}\ \emph {et~al.}(2005)\citenamefont
  {DiLuzio}, \citenamefont {Turner}, \citenamefont {Mayer}, \citenamefont
  {Garstecki}, \citenamefont {Weibel}, \citenamefont {Berg},\ and\
  \citenamefont {Whitesides}}]{DiLuzio2005}%
  \BibitemOpen
  \bibfield  {author} {\bibinfo {author} {\bibfnamefont {W.~R.}\ \bibnamefont
  {DiLuzio}}, \bibinfo {author} {\bibfnamefont {L.}~\bibnamefont {Turner}},
  \bibinfo {author} {\bibfnamefont {M.}~\bibnamefont {Mayer}}, \bibinfo
  {author} {\bibfnamefont {P.}~\bibnamefont {Garstecki}}, \bibinfo {author}
  {\bibfnamefont {D.~B.}\ \bibnamefont {Weibel}}, \bibinfo {author}
  {\bibfnamefont {H.~C.}\ \bibnamefont {Berg}}, \ and\ \bibinfo {author}
  {\bibfnamefont {G.~M.}\ \bibnamefont {Whitesides}},\ }\bibfield  {title}
  {\enquote {\bibinfo {title} {Escherichia coli swim on the right-hand side},}\
  }\href@noop {} {\bibfield  {journal} {\bibinfo  {journal} {Nature}\ }\textbf
  {\bibinfo {volume} {435}},\ \bibinfo {pages} {1271} (\bibinfo {year}
  {2005})}\BibitemShut {NoStop}%
\bibitem [{\citenamefont {Lauga}\ \emph {et~al.}(2006)\citenamefont {Lauga},
  \citenamefont {DiLuzio}, \citenamefont {Whitesides},\ and\ \citenamefont
  {Stone}}]{Lauga2006}%
  \BibitemOpen
  \bibfield  {author} {\bibinfo {author} {\bibfnamefont {E.}~\bibnamefont
  {Lauga}}, \bibinfo {author} {\bibfnamefont {W.~R.}\ \bibnamefont {DiLuzio}},
  \bibinfo {author} {\bibfnamefont {G.~M.}\ \bibnamefont {Whitesides}}, \ and\
  \bibinfo {author} {\bibfnamefont {H.~A.}\ \bibnamefont {Stone}},\ }\bibfield
  {title} {\enquote {\bibinfo {title} {Swimming in circles: motion of bacteria
  near solid boundaries},}\ }\href@noop {} {\bibfield  {journal} {\bibinfo
  {journal} {Biophys. J.}\ }\textbf {\bibinfo {volume} {90}},\ \bibinfo {pages}
  {400} (\bibinfo {year} {2006})}\BibitemShut {NoStop}%
\bibitem [{\citenamefont {K{\"u}mmel}\ \emph {et~al.}(2013)\citenamefont
  {K{\"u}mmel}, \citenamefont {ten Hagen}, \citenamefont {Wittkowski},
  \citenamefont {Buttinoni}, \citenamefont {Eichhorn}, \citenamefont {Volpe},
  \citenamefont {L{\"o}wen},\ and\ \citenamefont {Bechinger}}]{Kummel2013}%
  \BibitemOpen
  \bibfield  {author} {\bibinfo {author} {\bibfnamefont {F.}~\bibnamefont
  {K{\"u}mmel}}, \bibinfo {author} {\bibfnamefont {B.}~\bibnamefont {ten
  Hagen}}, \bibinfo {author} {\bibfnamefont {R.}~\bibnamefont {Wittkowski}},
  \bibinfo {author} {\bibfnamefont {I.}~\bibnamefont {Buttinoni}}, \bibinfo
  {author} {\bibfnamefont {R.}~\bibnamefont {Eichhorn}}, \bibinfo {author}
  {\bibfnamefont {G.}~\bibnamefont {Volpe}}, \bibinfo {author} {\bibfnamefont
  {H.}~\bibnamefont {L{\"o}wen}}, \ and\ \bibinfo {author} {\bibfnamefont
  {C.}~\bibnamefont {Bechinger}},\ }\bibfield  {title} {\enquote {\bibinfo
  {title} {Circular motion of asymmetric self-propelling particles},}\
  }\href@noop {} {\bibfield  {journal} {\bibinfo  {journal} {Phys. Rev. Lett.}\
  }\textbf {\bibinfo {volume} {110}},\ \bibinfo {pages} {198302} (\bibinfo
  {year} {2013})}\BibitemShut {NoStop}%
\bibitem [{\citenamefont {Wykes}\ \emph {et~al.}(2016)\citenamefont {Wykes},
  \citenamefont {Palacci}, \citenamefont {Adachi}, \citenamefont {Ristroph},
  \citenamefont {Zhong}, \citenamefont {Ward}, \citenamefont {Zhang},\ and\
  \citenamefont {Shelley}}]{Wykes2016}%
  \BibitemOpen
  \bibfield  {author} {\bibinfo {author} {\bibfnamefont {M.~S.~D.}\
  \bibnamefont {Wykes}}, \bibinfo {author} {\bibfnamefont {J.}~\bibnamefont
  {Palacci}}, \bibinfo {author} {\bibfnamefont {T.}~\bibnamefont {Adachi}},
  \bibinfo {author} {\bibfnamefont {L.}~\bibnamefont {Ristroph}}, \bibinfo
  {author} {\bibfnamefont {X.}~\bibnamefont {Zhong}}, \bibinfo {author}
  {\bibfnamefont {M.~D.}\ \bibnamefont {Ward}}, \bibinfo {author}
  {\bibfnamefont {J.}~\bibnamefont {Zhang}}, \ and\ \bibinfo {author}
  {\bibfnamefont {M.~J.}\ \bibnamefont {Shelley}},\ }\bibfield  {title}
  {\enquote {\bibinfo {title} {Dynamic self-assembly of microscale rotors and
  swimmers},}\ }\href@noop {} {\bibfield  {journal} {\bibinfo  {journal} {Soft
  Matter}\ }\textbf {\bibinfo {volume} {12}},\ \bibinfo {pages} {4584}
  (\bibinfo {year} {2016})}\BibitemShut {NoStop}%
\bibitem [{\citenamefont {Scholz}\ \emph {et~al.}(2018)\citenamefont {Scholz},
  \citenamefont {Jahanshahi}, \citenamefont {Ldov},\ and\ \citenamefont
  {L{\"o}wen}}]{Scholz2018}%
  \BibitemOpen
  \bibfield  {author} {\bibinfo {author} {\bibfnamefont {C.}~\bibnamefont
  {Scholz}}, \bibinfo {author} {\bibfnamefont {S.}~\bibnamefont {Jahanshahi}},
  \bibinfo {author} {\bibfnamefont {A.}~\bibnamefont {Ldov}}, \ and\ \bibinfo
  {author} {\bibfnamefont {H.}~\bibnamefont {L{\"o}wen}},\ }\bibfield  {title}
  {\enquote {\bibinfo {title} {Inertial delay of self-propelled particles},}\
  }\href@noop {} {\bibfield  {journal} {\bibinfo  {journal} {arXiv preprint
  arXiv:1807.04357}\ } (\bibinfo {year} {2018})}\BibitemShut {NoStop}%
\bibitem [{\citenamefont {Scholz}, \citenamefont {D'~Silva},\ and\
  \citenamefont {P{\"o}schel}(2016)}]{Scholz2016}%
  \BibitemOpen
  \bibfield  {author} {\bibinfo {author} {\bibfnamefont {C.}~\bibnamefont
  {Scholz}}, \bibinfo {author} {\bibfnamefont {S.}~\bibnamefont {D'~Silva}}, \
  and\ \bibinfo {author} {\bibfnamefont {T.}~\bibnamefont {P{\"o}schel}},\
  }\bibfield  {title} {\enquote {\bibinfo {title} {Ratcheting and tumbling
  motion of vibrots},}\ }\href@noop {} {\bibfield  {journal} {\bibinfo
  {journal} {New J. Physics}\ }\textbf {\bibinfo {volume} {18}},\ \bibinfo
  {pages} {123001} (\bibinfo {year} {2016})}\BibitemShut {NoStop}%
\bibitem [{\citenamefont {Mijalkov}\ and\ \citenamefont
  {Volpe}(2013)}]{Mijalkov2013}%
  \BibitemOpen
  \bibfield  {author} {\bibinfo {author} {\bibfnamefont {M.}~\bibnamefont
  {Mijalkov}}\ and\ \bibinfo {author} {\bibfnamefont {G.}~\bibnamefont
  {Volpe}},\ }\bibfield  {title} {\enquote {\bibinfo {title} {Sorting of chiral
  microswimmers},}\ }\href@noop {} {\bibfield  {journal} {\bibinfo  {journal}
  {Soft Matter}\ }\textbf {\bibinfo {volume} {9}},\ \bibinfo {pages} {6376}
  (\bibinfo {year} {2013})}\BibitemShut {NoStop}%
\bibitem [{\citenamefont {Chen}\ and\ \citenamefont {Ai}(2015)}]{Chen2015}%
  \BibitemOpen
  \bibfield  {author} {\bibinfo {author} {\bibfnamefont {Q.}~\bibnamefont
  {Chen}}\ and\ \bibinfo {author} {\bibfnamefont {B.-q.}\ \bibnamefont {Ai}},\
  }\bibfield  {title} {\enquote {\bibinfo {title} {Sorting of chiral active
  particles driven by rotary obstacles},}\ }\href@noop {} {\bibfield  {journal}
  {\bibinfo  {journal} {J. Chem. Phys.}\ }\textbf {\bibinfo {volume} {143}},\
  \bibinfo {pages} {09B612\_1} (\bibinfo {year} {2015})}\BibitemShut {NoStop}%
\bibitem [{\citenamefont {Levis}, \citenamefont {Pagonabarraga},\ and\
  \citenamefont {Liebchen}(2018)}]{Levis2018}%
  \BibitemOpen
  \bibfield  {author} {\bibinfo {author} {\bibfnamefont {D.}~\bibnamefont
  {Levis}}, \bibinfo {author} {\bibfnamefont {I.}~\bibnamefont
  {Pagonabarraga}}, \ and\ \bibinfo {author} {\bibfnamefont {B.}~\bibnamefont
  {Liebchen}},\ }\bibfield  {title} {\enquote {\bibinfo {title} {Activity
  induced synchronization},}\ }\href@noop {} {\bibfield  {journal} {\bibinfo
  {journal} {arXiv preprint arXiv:1802.02371}\ } (\bibinfo {year}
  {2018})}\BibitemShut {NoStop}%
\bibitem [{\citenamefont {Chepizhko}\ and\ \citenamefont
  {Franosch}(2018)}]{Chepi}%
  \BibitemOpen
  \bibfield  {author} {\bibinfo {author} {\bibfnamefont {O.}~\bibnamefont
  {Chepizhko}}\ and\ \bibinfo {author} {\bibfnamefont {T.}~\bibnamefont
  {Franosch}},\ }\bibfield  {title} {\enquote {\bibinfo {title} {Ideal circle
  microswimmers in crowded media},}\ }\href@noop {} {\bibfield  {journal}
  {\bibinfo  {journal} {Soft Matter}\ ,\ \bibinfo {pages} {Accepted
  Manuscript}} (\bibinfo {year} {2018})}\BibitemShut {NoStop}%
\bibitem [{\citenamefont {Kaiser}\ and\ \citenamefont
  {L{\"o}wen}(2013)}]{Kaiser2013}%
  \BibitemOpen
  \bibfield  {author} {\bibinfo {author} {\bibfnamefont {A.}~\bibnamefont
  {Kaiser}}\ and\ \bibinfo {author} {\bibfnamefont {H.}~\bibnamefont
  {L{\"o}wen}},\ }\bibfield  {title} {\enquote {\bibinfo {title} {Vortex arrays
  as emergent collective phenomena for circle swimmers},}\ }\href@noop {}
  {\bibfield  {journal} {\bibinfo  {journal} {Phys. Rev. E}\ }\textbf {\bibinfo
  {volume} {87}},\ \bibinfo {pages} {032712} (\bibinfo {year}
  {2013})}\BibitemShut {NoStop}%
\bibitem [{\citenamefont {Denk}\ \emph {et~al.}(2016)\citenamefont {Denk},
  \citenamefont {Huber}, \citenamefont {Reithmann},\ and\ \citenamefont
  {Frey}}]{Denk2016}%
  \BibitemOpen
  \bibfield  {author} {\bibinfo {author} {\bibfnamefont {J.}~\bibnamefont
  {Denk}}, \bibinfo {author} {\bibfnamefont {L.}~\bibnamefont {Huber}},
  \bibinfo {author} {\bibfnamefont {E.}~\bibnamefont {Reithmann}}, \ and\
  \bibinfo {author} {\bibfnamefont {E.}~\bibnamefont {Frey}},\ }\bibfield
  {title} {\enquote {\bibinfo {title} {Active curved polymers form vortex
  patterns on membranes},}\ }\href@noop {} {\bibfield  {journal} {\bibinfo
  {journal} {Phys. Rev. Lett.}\ }\textbf {\bibinfo {volume} {116}},\ \bibinfo
  {pages} {178301} (\bibinfo {year} {2016})}\BibitemShut {NoStop}%
\bibitem [{\citenamefont {Liebchen}\ and\ \citenamefont
  {Levis}(2017)}]{Liebchen2017}%
  \BibitemOpen
  \bibfield  {author} {\bibinfo {author} {\bibfnamefont {B.}~\bibnamefont
  {Liebchen}}\ and\ \bibinfo {author} {\bibfnamefont {D.}~\bibnamefont
  {Levis}},\ }\bibfield  {title} {\enquote {\bibinfo {title} {Collective
  behavior of chiral active matter: pattern formation and enhanced flocking},}\
  }\href@noop {} {\bibfield  {journal} {\bibinfo  {journal} {Phys. Rev. Lett.}\
  }\textbf {\bibinfo {volume} {119}},\ \bibinfo {pages} {058002} (\bibinfo
  {year} {2017})}\BibitemShut {NoStop}%
\bibitem [{\citenamefont {Loose}\ and\ \citenamefont
  {Mitchison}(2014)}]{Loose2014}%
  \BibitemOpen
  \bibfield  {author} {\bibinfo {author} {\bibfnamefont {M.}~\bibnamefont
  {Loose}}\ and\ \bibinfo {author} {\bibfnamefont {T.~J.}\ \bibnamefont
  {Mitchison}},\ }\bibfield  {title} {\enquote {\bibinfo {title} {The bacterial
  cell division proteins ftsa and ftsz self-organize into dynamic cytoskeletal
  patterns},}\ }\href@noop {} {\bibfield  {journal} {\bibinfo  {journal} {Nat.
  Cell. Biol.}\ }\textbf {\bibinfo {volume} {16}},\ \bibinfo {pages} {38}
  (\bibinfo {year} {2014})}\BibitemShut {NoStop}%
\bibitem [{\citenamefont {Levis}\ and\ \citenamefont
  {Liebchen}(2018)}]{Levis2018micro}%
  \BibitemOpen
  \bibfield  {author} {\bibinfo {author} {\bibfnamefont {D.}~\bibnamefont
  {Levis}}\ and\ \bibinfo {author} {\bibfnamefont {B.}~\bibnamefont
  {Liebchen}},\ }\bibfield  {title} {\enquote {\bibinfo {title} {Micro-flock
  patterns and macro-clusters in chiral active brownian disks},}\ }\href@noop
  {} {\bibfield  {journal} {\bibinfo  {journal} {J. Phys. Cond. Matter}\
  }\textbf {\bibinfo {volume} {30}},\ \bibinfo {pages} {084001} (\bibinfo
  {year} {2018})}\BibitemShut {NoStop}%
\bibitem [{\citenamefont {Lei}, \citenamefont {Ciamarra},\ and\ \citenamefont
  {Ni}(2018)}]{Lei2018}%
  \BibitemOpen
  \bibfield  {author} {\bibinfo {author} {\bibfnamefont {Q.-l.}\ \bibnamefont
  {Lei}}, \bibinfo {author} {\bibfnamefont {M.~P.}\ \bibnamefont {Ciamarra}}, \
  and\ \bibinfo {author} {\bibfnamefont {R.}~\bibnamefont {Ni}},\ }\bibfield
  {title} {\enquote {\bibinfo {title} {Non-equilibrium strong hyperuniform
  fluids of athermal active circle swimmers with giant local fluctuations},}\
  }\href@noop {} {\bibfield  {journal} {\bibinfo  {journal} {arXiv preprint
  arXiv:1802.03682}\ } (\bibinfo {year} {2018})}\BibitemShut {NoStop}%
\bibitem [{\citenamefont {Liao}\ and\ \citenamefont {Klapp}(2018)}]{Liao2018}%
  \BibitemOpen
  \bibfield  {author} {\bibinfo {author} {\bibfnamefont {G.-J.}\ \bibnamefont
  {Liao}}\ and\ \bibinfo {author} {\bibfnamefont {S.~H.}\ \bibnamefont
  {Klapp}},\ }\bibfield  {title} {\enquote {\bibinfo {title} {Clustering and
  phase separation of circle swimmers dispersed in a monolayer},}\ }\href@noop
  {} {\bibfield  {journal} {\bibinfo  {journal} {Soft Matter}\ }\textbf
  {\bibinfo {volume} {14}},\ \bibinfo {pages} {7873} (\bibinfo {year}
  {2018})}\BibitemShut {NoStop}%
\bibitem [{\citenamefont {Reichhardt}\ and\ \citenamefont
  {Reichhardt}(2018)}]{Reichhardt2018}%
  \BibitemOpen
  \bibfield  {author} {\bibinfo {author} {\bibfnamefont {D.}~\bibnamefont
  {Reichhardt}}\ and\ \bibinfo {author} {\bibfnamefont {C.~J.~O.}\ \bibnamefont
  {Reichhardt}},\ }\bibfield  {title} {\enquote {\bibinfo {title}
  {Reversibility, pattern formation and edge transport in active chiral and
  passive disk mixtures},}\ }\href@noop {} {\bibfield  {journal} {\bibinfo
  {journal} {arXiv preprint arXiv:1812.04150}\ } (\bibinfo {year}
  {2018})}\BibitemShut {NoStop}%
\bibitem [{\citenamefont {Ai}, \citenamefont {Shao},\ and\ \citenamefont
  {Zhong}(2018)}]{Ai2018}%
  \BibitemOpen
  \bibfield  {author} {\bibinfo {author} {\bibfnamefont {B.-q.}\ \bibnamefont
  {Ai}}, \bibinfo {author} {\bibfnamefont {Z.-g.}\ \bibnamefont {Shao}}, \ and\
  \bibinfo {author} {\bibfnamefont {W.-r.}\ \bibnamefont {Zhong}},\ }\bibfield
  {title} {\enquote {\bibinfo {title} {Mixing and demixing of binary mixtures
  of polar chiral active particles},}\ }\href@noop {} {\bibfield  {journal}
  {\bibinfo  {journal} {Soft matter}\ }\textbf {\bibinfo {volume} {14}},\
  \bibinfo {pages} {4388} (\bibinfo {year} {2018})}\BibitemShut {NoStop}%
\bibitem [{\citenamefont {van Teeffelen}\ and\ \citenamefont
  {L{\"o}wen}(2008)}]{vanTeeffelen2008}%
  \BibitemOpen
  \bibfield  {author} {\bibinfo {author} {\bibfnamefont {S.}~\bibnamefont {van
  Teeffelen}}\ and\ \bibinfo {author} {\bibfnamefont {H.}~\bibnamefont
  {L{\"o}wen}},\ }\bibfield  {title} {\enquote {\bibinfo {title} {Dynamics of a
  brownian circle swimmer},}\ }\href@noop {} {\bibfield  {journal} {\bibinfo
  {journal} {Phys. Rev. E}\ }\textbf {\bibinfo {volume} {78}},\ \bibinfo
  {pages} {020101} (\bibinfo {year} {2008})}\BibitemShut {NoStop}%
\bibitem [{\citenamefont {Sevilla}(2016)}]{Sevilla2016}%
  \BibitemOpen
  \bibfield  {author} {\bibinfo {author} {\bibfnamefont {F.~J.}\ \bibnamefont
  {Sevilla}},\ }\bibfield  {title} {\enquote {\bibinfo {title} {Diffusion of
  active chiral particles},}\ }\href@noop {} {\bibfield  {journal} {\bibinfo
  {journal} {Phys. Rev. E}\ }\textbf {\bibinfo {volume} {94}},\ \bibinfo
  {pages} {062120} (\bibinfo {year} {2016})}\BibitemShut {NoStop}%
\bibitem [{\citenamefont {Vicsek}\ and\ \citenamefont
  {Zafeiris}(2012)}]{VicsekRev}%
  \BibitemOpen
  \bibfield  {author} {\bibinfo {author} {\bibfnamefont {T.}~\bibnamefont
  {Vicsek}}\ and\ \bibinfo {author} {\bibfnamefont {A.}~\bibnamefont
  {Zafeiris}},\ }\bibfield  {title} {\enquote {\bibinfo {title} {Collective
  motion},}\ }\href@noop {} {\bibfield  {journal} {\bibinfo  {journal} {Phys.
  Rep.}\ }\textbf {\bibinfo {volume} {517}},\ \bibinfo {pages} {71} (\bibinfo
  {year} {2012})}\BibitemShut {NoStop}%
\bibitem [{\citenamefont {Toner}\ and\ \citenamefont {Tu}(1995)}]{Toner1995}%
  \BibitemOpen
  \bibfield  {author} {\bibinfo {author} {\bibfnamefont {J.}~\bibnamefont
  {Toner}}\ and\ \bibinfo {author} {\bibfnamefont {Y.}~\bibnamefont {Tu}},\
  }\bibfield  {title} {\enquote {\bibinfo {title} {Long-range order in a
  two-dimensional dynamical xy model: how birds fly together},}\ }\href@noop {}
  {\bibfield  {journal} {\bibinfo  {journal} {Phys. Rev. Lett.}\ }\textbf
  {\bibinfo {volume} {75}},\ \bibinfo {pages} {4326} (\bibinfo {year}
  {1995})}\BibitemShut {NoStop}%
\bibitem [{\citenamefont {Acebr{\'o}n}\ \emph {et~al.}(2005)\citenamefont
  {Acebr{\'o}n}, \citenamefont {Bonilla}, \citenamefont {P{\'e}rez-Vicente},
  \citenamefont {Ritort},\ and\ \citenamefont {Spigler}}]{Acebron2005}%
  \BibitemOpen
  \bibfield  {author} {\bibinfo {author} {\bibfnamefont {J.~A.}\ \bibnamefont
  {Acebr{\'o}n}}, \bibinfo {author} {\bibfnamefont {L.~L.}\ \bibnamefont
  {Bonilla}}, \bibinfo {author} {\bibfnamefont {C.~J.}\ \bibnamefont
  {P{\'e}rez-Vicente}}, \bibinfo {author} {\bibfnamefont {F.}~\bibnamefont
  {Ritort}}, \ and\ \bibinfo {author} {\bibfnamefont {R.}~\bibnamefont
  {Spigler}},\ }\bibfield  {title} {\enquote {\bibinfo {title} {The kuramoto
  model: A simple paradigm for synchronization phenomena},}\ }\href@noop {}
  {\bibfield  {journal} {\bibinfo  {journal} {Rev. Mod. Phys.}\ }\textbf
  {\bibinfo {volume} {77}},\ \bibinfo {pages} {137} (\bibinfo {year}
  {2005})}\BibitemShut {NoStop}%
\bibitem [{\citenamefont {Berezinskii}(1971)}]{berezinskii1971}%
  \BibitemOpen
  \bibfield  {author} {\bibinfo {author} {\bibfnamefont {V.}~\bibnamefont
  {Berezinskii}},\ }\bibfield  {title} {\enquote {\bibinfo {title} {Destruction
  of long-range order in one-dimensional and two-dimensional systems having a
  continuous symmetry group i. classical systems},}\ }\href@noop {} {\bibfield
  {journal} {\bibinfo  {journal} {Sov. Phys. JETP}\ }\textbf {\bibinfo {volume}
  {32}},\ \bibinfo {pages} {493--500} (\bibinfo {year} {1971})}\BibitemShut
  {NoStop}%
\bibitem [{\citenamefont {Daido}(1988)}]{Daido1988}%
  \BibitemOpen
  \bibfield  {author} {\bibinfo {author} {\bibfnamefont {H.}~\bibnamefont
  {Daido}},\ }\bibfield  {title} {\enquote {\bibinfo {title} {Lower critical
  dimension for populations of oscillators with randomly distributed
  frequencies: a renormalization-group analysis},}\ }\href@noop {} {\bibfield
  {journal} {\bibinfo  {journal} {Phys. Rev. Lett.}\ }\textbf {\bibinfo
  {volume} {61}},\ \bibinfo {pages} {231} (\bibinfo {year} {1988})}\BibitemShut
  {NoStop}%
\bibitem [{\citenamefont {Mermin}\ and\ \citenamefont
  {Wagner}(1966)}]{Mermin1966}%
  \BibitemOpen
  \bibfield  {author} {\bibinfo {author} {\bibfnamefont {N.~D.}\ \bibnamefont
  {Mermin}}\ and\ \bibinfo {author} {\bibfnamefont {H.}~\bibnamefont
  {Wagner}},\ }\bibfield  {title} {\enquote {\bibinfo {title} {Absence of
  ferromagnetism or antiferromagnetism in one-or two-dimensional isotropic
  heisenberg models},}\ }\href@noop {} {\bibfield  {journal} {\bibinfo
  {journal} {Phys. Rev. Lett.}\ }\textbf {\bibinfo {volume} {17}},\ \bibinfo
  {pages} {1133} (\bibinfo {year} {1966})}\BibitemShut {NoStop}%
\bibitem [{\citenamefont {Mart{\'\i}n-G{\'o}mez}\ \emph
  {et~al.}(2018)\citenamefont {Mart{\'\i}n-G{\'o}mez}, \citenamefont {Levis},
  \citenamefont {D{\'\i}az-Guilera},\ and\ \citenamefont
  {Pagonabarraga}}]{Aitor}%
  \BibitemOpen
  \bibfield  {author} {\bibinfo {author} {\bibfnamefont {A.}~\bibnamefont
  {Mart{\'\i}n-G{\'o}mez}}, \bibinfo {author} {\bibfnamefont {D.}~\bibnamefont
  {Levis}}, \bibinfo {author} {\bibfnamefont {A.}~\bibnamefont
  {D{\'\i}az-Guilera}}, \ and\ \bibinfo {author} {\bibfnamefont
  {I.}~\bibnamefont {Pagonabarraga}},\ }\bibfield  {title} {\enquote {\bibinfo
  {title} {Collective motion of active brownian particles with polar
  alignment},}\ }\href@noop {} {\bibfield  {journal} {\bibinfo  {journal} {Soft
  Matter}\ }\textbf {\bibinfo {volume} {14}},\ \bibinfo {pages} {2610}
  (\bibinfo {year} {2018})}\BibitemShut {NoStop}%
\bibitem [{\citenamefont {Narayan}, \citenamefont {Ramaswamy},\ and\
  \citenamefont {Menon}(2007)}]{Narayan2007}%
  \BibitemOpen
  \bibfield  {author} {\bibinfo {author} {\bibfnamefont {V.}~\bibnamefont
  {Narayan}}, \bibinfo {author} {\bibfnamefont {S.}~\bibnamefont {Ramaswamy}},
  \ and\ \bibinfo {author} {\bibfnamefont {N.}~\bibnamefont {Menon}},\
  }\bibfield  {title} {\enquote {\bibinfo {title} {Long-lived giant number
  fluctuations in a swarming granular nematic},}\ }\href@noop {} {\bibfield
  {journal} {\bibinfo  {journal} {Science}\ }\textbf {\bibinfo {volume}
  {317}},\ \bibinfo {pages} {105} (\bibinfo {year} {2007})}\BibitemShut
  {NoStop}%
\bibitem [{\citenamefont {Chat{\'e}}\ \emph {et~al.}(2008)\citenamefont
  {Chat{\'e}}, \citenamefont {Ginelli}, \citenamefont {Gr{\'e}goire},\ and\
  \citenamefont {Raynaud}}]{Chate2008}%
  \BibitemOpen
  \bibfield  {author} {\bibinfo {author} {\bibfnamefont {H.}~\bibnamefont
  {Chat{\'e}}}, \bibinfo {author} {\bibfnamefont {F.}~\bibnamefont {Ginelli}},
  \bibinfo {author} {\bibfnamefont {G.}~\bibnamefont {Gr{\'e}goire}}, \ and\
  \bibinfo {author} {\bibfnamefont {F.}~\bibnamefont {Raynaud}},\ }\bibfield
  {title} {\enquote {\bibinfo {title} {Collective motion of self-propelled
  particles interacting without cohesion},}\ }\href@noop {} {\bibfield
  {journal} {\bibinfo  {journal} {Phys. Rev. E}\ }\textbf {\bibinfo {volume}
  {77}},\ \bibinfo {pages} {046113} (\bibinfo {year} {2008})}\BibitemShut
  {NoStop}%
\bibitem [{SM()}]{SM}%
  \BibitemOpen
  \href@noop {} {\bibinfo  {journal} {See Supplementary Material at doi:...}\
  }\BibitemShut {NoStop}%
\bibitem [{\citenamefont {Hoell}, \citenamefont {L{\"o}wen},\ and\
  \citenamefont {Menzel}(2017)}]{Hoell2017}%
  \BibitemOpen
\bibfield  {journal} {  }\bibfield  {author} {\bibinfo {author} {\bibfnamefont
  {C.}~\bibnamefont {Hoell}}, \bibinfo {author} {\bibfnamefont
  {H.}~\bibnamefont {L{\"o}wen}}, \ and\ \bibinfo {author} {\bibfnamefont
  {A.~M.}\ \bibnamefont {Menzel}},\ }\bibfield  {title} {\enquote {\bibinfo
  {title} {Dynamical density functional theory for circle swimmers},}\
  }\href@noop {} {\bibfield  {journal} {\bibinfo  {journal} {New J. Phys.}\
  }\textbf {\bibinfo {volume} {19}},\ \bibinfo {pages} {125004} (\bibinfo
  {year} {2017})}\BibitemShut {NoStop}%
\bibitem [{\citenamefont {Farrell}\ \emph {et~al.}(2012)\citenamefont
  {Farrell}, \citenamefont {Marchetti}, \citenamefont {Marenduzzo},\ and\
  \citenamefont {Tailleur}}]{Farrell2012}%
  \BibitemOpen
  \bibfield  {author} {\bibinfo {author} {\bibfnamefont {F.~D.~C.}\
  \bibnamefont {Farrell}}, \bibinfo {author} {\bibfnamefont {M.~C.}\
  \bibnamefont {Marchetti}}, \bibinfo {author} {\bibfnamefont {D.}~\bibnamefont
  {Marenduzzo}}, \ and\ \bibinfo {author} {\bibfnamefont {J.}~\bibnamefont
  {Tailleur}},\ }\bibfield  {title} {\enquote {\bibinfo {title} {Pattern
  formation in self-propelled particles with density-dependent motility},}\
  }\href@noop {} {\bibfield  {journal} {\bibinfo  {journal} {Phys. Rev. Lett.}\
  }\textbf {\bibinfo {volume} {108}},\ \bibinfo {pages} {248101} (\bibinfo
  {year} {2012})}\BibitemShut {NoStop}%
\end{thebibliography}%

\end{document}